\documentclass[prb,twocolumn,superscriptaddress,showpacs,amsmath,amssymb]{revtex4-1}

\usepackage{graphicx}
\usepackage{latexsym}
\usepackage{amsmath}
\usepackage{amssymb}
\usepackage{amsfonts}
\usepackage{changes}
\usepackage{color}
\usepackage{bm}
\usepackage{verbatim}
\usepackage{pagecolor,lipsum}
\usepackage{lineno}
\usepackage[version=3]{mhchem}

\usepackage{framed}
\definecolor{shadecolor}{RGB}{222,222,221}
\begin{document}

\title{Proximity effect in superconductor/antiferromagnet hybrids: N\'eel triplets and impurity suppression of superconductivity}

\author{G.A. Bobkov}
\affiliation{Moscow Institute of Physics and Technology, Dolgoprudny, 141700 Russia}

\author{I. V. Bobkova}
\affiliation{Moscow Institute of Physics and Technology, Dolgoprudny, 141700 Russia}
\affiliation{National Research University Higher School of Economics, Moscow, 101000 Russia}

\author{A. M. Bobkov}
\affiliation{Moscow Institute of Physics and Technology, Dolgoprudny, 141700 Russia}

\date{\today}


\begin{abstract}
Two possible
physical mechanisms of superconductivity suppression at superconductor/antiferromagnet (S/AF) interfaces, which work even for interfaces with compensated antiferromagnets, were reported.  
One of them suggests that the N\'eel order of the AF
induces rapidly oscillating spin-triplet correlations in the S layer. They are called N\'eel triplets, and they suppress singlet superconductivity.  Nonmagnetic
disorder destroys this type of triplet correlations. As a result, the critical temperature of the S/AF bilayer grows with impurity strength. The second mechanism, on the contrary, suggests that nonmagnetic impurity scattering suppresses superconductivity in
S/AF hybrids. The predictions were made in the framework of two different quasiclassical approaches [G. A. Bobkov {\it et~al.} Phys. Rev. B 106, 144512 (2022) and E. H. Fyhn {\it et~al.}
arXiv:2210.09325]. Here we suggest the unified theory of the proximity effect in thin-film S/AF structures, which incorporates both pictures as limiting cases, and we study the proximity effect at S/AF interfaces for arbitrary impurity strength, chemical potential, and the value of the N\'eel exchange field. 
\end{abstract}

  \maketitle
 
\section{Introduction}

It is well established that nanoscale regions near the superconductor/ferromagnet (S/F) interfaces and artificial thin-film S/F systems bring new fundamental physics and applications, which are not possible in these materials separately. In particular, odd-frequency superconductivity \cite{Linder2019} can be produced, and spinful triplet Cooper pairs \cite{Bergeret2005,Eschrig2015,Linder2015}  and  composite magnetic excitations magnon-Cooparons\cite{Bobkova2022}, which carry dissipationless spin-currents, can appear and be exploited for spintronics applications. Zeeman-split superconductivity at S/F interfaces can lead to a giant spin-dependent Seebeck effect \cite{Bergeret2018,Heikkila2019} and highly efficient motion of magnetic defects \cite{Bobkova2021}. All these effects are due to the fact that interfacial
exchange interaction with a ferromagnet causes spin-singlet pairs to be converted into
their spin-triplet counterparts \cite{Buzdin2005,Bergeret2018,Heikkila2019}, and it leads to the occurrence of spin-polarized quasiparticles.  

On the other hand, antiferromagnets (AFs) are being actively studied as alternatives to ferromagnets for spintronics applications due to their robustness against perturbation by magnetic fields, the absence of parasitic stray fields, and their ultrafast dynamics \cite{Baltz2018,Jungwirth2016,Brataas2020}. Superconductor/antiferromagnet (S/AF) heterostructures have been studied both theoretically and experimentally, but much less than S/F hybrids. It has been shown that if the S/AF interface possesses nonzero net magnetization (uncompensated interface), it also induces Zeeman splitting in the adjacent superconductor \cite{Kamra2018}. This leads to similar fundamental physics and  prospects for applications, as for S/F structures. In particular,  composite magnonic excitations \cite{Bobkov2022},  the giant spin-dependent Seebeck effect \cite{Bobkov2021}, atomic thickness $0-\pi$ transitions \cite{Andersen2006,Enoksen2013}, and magnetoelectric effects in Josephson junctions via uncompensated antiferromagnets \cite{Rabinovich2019,Falch2022} have been predicted. It has also been shown that topological superconductivity should occur in S/AF hybrids in the presence of spin-orbit coupling \cite{Lado2018}.

The physics of the proximity effect at the {\it compensated}  S/AF interface, i.e., if the magnetic moment of the interface is zero, is less clear. Due to the zero net resulting magnetization, such an interface was expected to experience no net spin-splitting or reduction in critical temperature \cite{Hauser1966,Kamra2018}. Nevertheless, unconventional Andreev reflection and bound
states at such S/AF interfaces have been predicted \cite{Bobkova2005,Andersen2005}. It was further predicted that the unconventional Andreev reflection results in qualitatively different transport
properties of S/AF junctions from those of superconductor/normal metal (S/N) and S/F junctions \cite{Jakobsen2020}. Also the spin-active behavior of S/AF interfaces has been demonstrated in AF/S/F heterostructures \cite{Johnsen2021}.
Several experiments have found that AFs suppress the critical temperature of an S layer \cite{Bell2003,Hubener2002,Wu2013,Seeger2021}, despite the absence of the net spin-splitting. In some cases, the effect has been comparable to, or even larger than, that induced by a ferromagnet layer \cite{Wu2013}. All this means that the proximity effect at the compensated S/AF interfaces is nontrivial and is not reduced to the well-known proximity physics of S/N interfaces. 

Several proposals have been suggested in the literature to explain this superconductivity suppression at S/AF interfaces. It could be, at least partially, due to  the well-known physical mechanisms such as the finite spin-splitting coming from uncompensated interfaces \cite{Bell2003}, the possible infusion of magnetic impurities into the superconductor during sample preparation \cite{Wu2013},  or the complex spin structure of the antiferromagnetic materials used in the experiments \cite{Hubener2002}. However, two other possible physical mechanisms, unique for the proximity effect at the S/AF interface, have been reported\cite{Bobkov2022_Neel,Fyhn2022_1}. 

One of them suggests that the N\'eel order of the AF induces spin-triplet correlations in the S layer\cite{Bobkov2022_Neel}. Their amplitude flips sign from one lattice site to the next, just like the N\'eel spin order in the AF. Thus, they were called N\'eel triplet Cooper pairs. The formation of such N\'eel spin triplets comes at the expense of destroying the spin-singlet correlation, which reduces the S critical temperature.  Nonmagnetic disorder destroys the N\'eel spin-triplet correlations and diminishes the proximity-induced suppression of superconductivity in the S/AF bilayer. Thus, this mechanism weakens superconductivity strongly 
for cleaner systems.  It is in sharp contrast with
the behavior of a thin-film F/S bilayer, where the degree of suppression is not sensitive to the {\it nonmagnetic} impurity concentration, and the {\it magnetic} impurities suppress the superconductivity \cite{Abrikosov1961}.    

\begin{figure}[tb]
	\begin{center}
		\includegraphics[width=85mm]{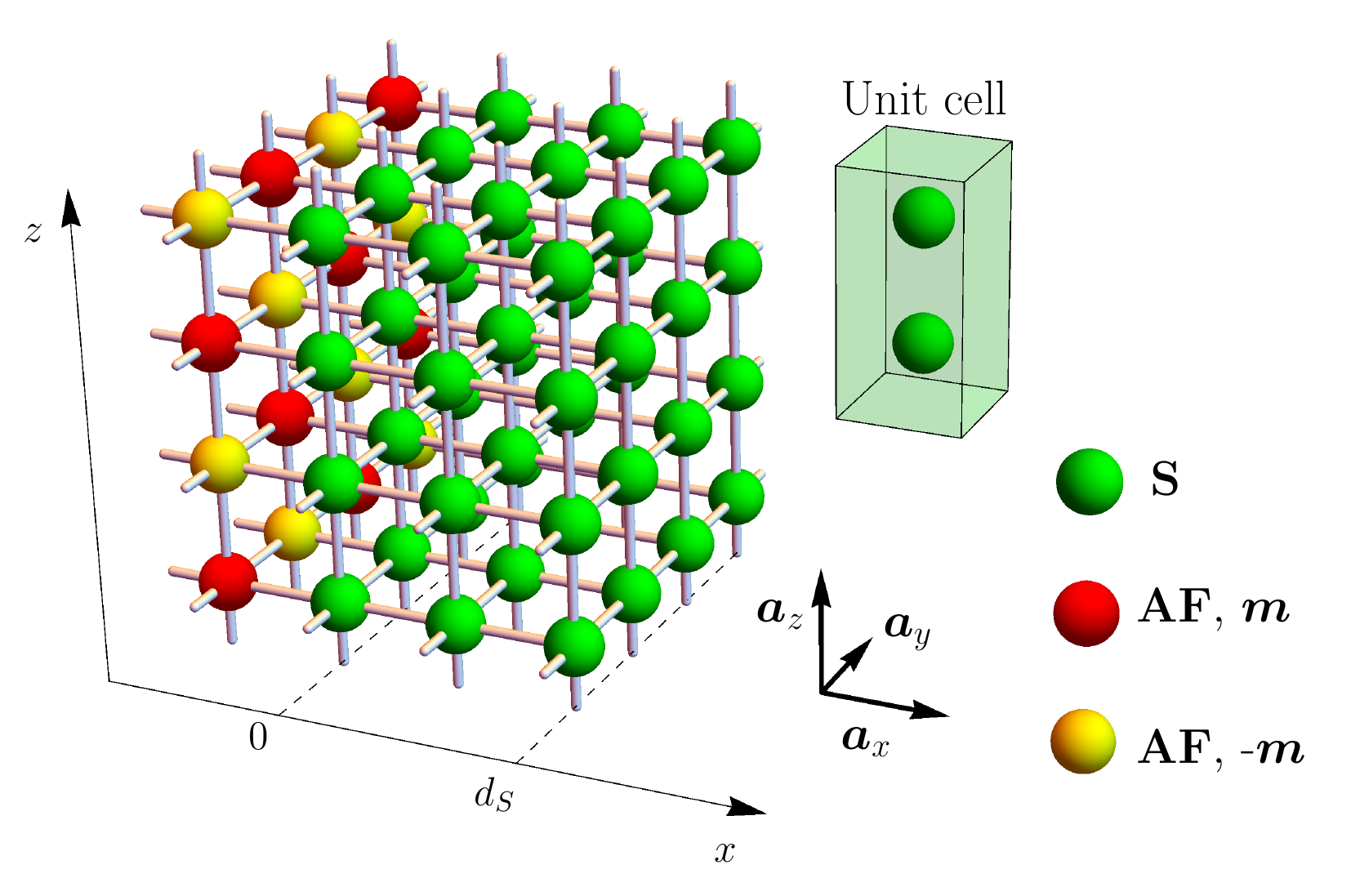}
		\caption{Sketch of the thin-film S/AF bilayer. The antiferromagnet occupies sites at $x<0$. The staggered magnetism is described by $\bm m^{A(B)} = + (-) \bm m$. The AF sites belonging to A and B sublattices are shown in red and yellow, respectively.  Only one antiferromagnetic layer is shown in the picture, but the AF is assumed to be semi-infinite. The AF is an insulator, therefore the electronic Green's functions are only calculated in the superconductor.  The superconducting sites are green.  The unit cell (taken in the superconductor) with two sites in it, belonging to A and B sublattices, is also shown. The S layer has a finite thickness $d_S$.}
        \label{fig:setup}
	\end{center}
\end{figure}

On the contrary, the second mechanism suggests that nonmagnetic impurity scattering suppresses superconductivity in S/AF hybrids.  The reason is that the sublattice-spin coupling in
the antiferromagnet gives an effective magnetic component to the non-magnetic impurities. The amplitude of the wave function of electrons is different for A and B sublattices because for an electron with spin up it is energetically favorable to be localized on a B sublattice, and for an electron with spin down it is energetically favorable to be localized on the A sublattice. The mechanism is similar to that discussed previously for antiferromagnetic superconductors \cite{Buzdin1986}. 

At first glance, two physical scenarios, described in two previous paragraphs, contradict each other and make opposite conclusions on the proximity effect in S/AF hybrid structures both in the clean limit and in the presence of disorder. Here we suggest the unified theory of the proximity effect in thin-film S/AF structures, which incorporates both pictures described above as limiting cases. The critical temperature of the S/AF thin-film bilayer is studied for an arbitrary concentration of impurities, values of the effective N\'eel exchange field $h$, induced in the superconductor by the AF, and the chemical potential of the superconductor $\mu$. It is shown that in general, both mechanisms of the superconductivity suppression by proximity to the AF, namely, the N\'eel triplets and the suppression by nonmagnetic impurities, exist. The key parameter controlling the amplitude of the N\'eel triplets is the ratio $h/\mu$.  The N\'eel triplets are always the dominating depairing factor in the clean case or if the disorder is weak. But for intermediate to strong disorder strength, there are two possible limiting scenarios, described in Refs.~\onlinecite{Bobkov2022_Neel,Fyhn2022_1}. The physics is controlled by the chemical potential ($\mu=0$ corresponds to half-filling of the conduction band): the superconductivity suppression is dominated by the N\'eel triplets at $\mu \lesssim T_c$, where $T_c$ is the critical temperature of the superconductor, and, consequently, the suppression is stronger for cleaner samples. On the contrary, if $\mu \gg T_c$ the superconductivity suppression is dominated by the nonmagnetic disorder. For intermediate values of $\mu$, a crossover between the regimes occurs. 

The paper is organized as follows. In Sec.~\ref{Gorkov} we describe the formalism of two-sublattice Gor'kov Green's functions beyond the quasiclassical approximation, which we exploit to develop the unified theory of the proximity effect in S/AF thin-film hybrids. In Sec.~\ref{results} the critical temperature of the structure is studied as a function of all the relevant parameters, and the crossover between opposite regimes of the response of the S/AF structure on the increase of the disorder strength is demonstrated. In Sec.~\ref{conclusions} we present our conclusions.

\section{Gor'kov equations}
\label{Gorkov}

We consider a thin-film S/AF bilayer, where the antiferromagnet is assumed to be an insulator, see Fig.~\ref{fig:setup}. The system is described in a two-sublattice framework. The unit cell with two sites in it, belonging to  $A$ and $B$ sublattices, is introduced as shown in Fig.~\ref{fig:setup}. In the framework of this two-sublattice approach the unit cells as a whole are marked by radius-vector $\bm i$. Then the staggered magnetism is described by $\bm m_{\bm i}^{A(B)} = + (-) \bm m_{\bm i}$, where $\bm m_{\bm i}$ is the local magnetic moment at site $A$ of the unit cell with the radius vector $\bm i$ in the AF. The influence of the antiferromagnetic insulator on the superconductor is described by the  exchange field $\bm h_{\bm i}^{A(B)} = \bm h_0 (-\bm h_0) \delta_{i_x,0}$, where the $\delta$-symbol means that the exchange field is only nonzero at the S/AF interface sites corresponding to $i_x = 0$ \cite{Kamra2018}. We assume that the interface is fully compensated, that is, the interface exchange field is staggered with zero average value. The superconductor S is described by the Hamiltonian: 
\begin{align}
\hat H= - t \sum \limits_{\langle \bm{i}\bm{j}\nu \bar \nu\rangle ,\sigma} \hat \psi_{\bm{i} \sigma}^{\nu\dagger} \hat \psi_{\bm{j} \sigma}^{\bar \nu} + \sum \limits_{\bm{i},\nu } (\Delta_{\bm{i}}^\nu \hat \psi_{\bm{i}\uparrow}^{\nu\dagger} \hat \psi_{\bm{i}\downarrow}^{\nu\dagger} + H.c.) - \nonumber \\
\mu \sum \limits_{\bm{i} \nu, \sigma} \hat n_{\bm{i}\sigma}^\nu  
+ \sum \limits_{\bm{i} \nu,\alpha \beta} \hat \psi_{\bm{i}\alpha}^{\nu \dagger} (\bm{h}_{\bm{i}}^\nu \bm{\sigma})_{\alpha \beta} \hat \psi_{\bm{i}\beta}^\nu + \sum \limits_{\bm{i}\nu,\sigma} V_i^\nu \hat n_{\bm{i}\sigma}^\nu,
\label{ham_2}
\end{align}
where $\nu=A,B$ is the sublattice index, $\bar \nu = A(B)$ if $\nu=B(A)$ means that the corresponding quantity belongs to the opposite sublattice, $\langle \bm i \bm j \nu \bar \nu\rangle $ means summation over the nearest neighbors, and $\hat{\psi}_{\bm i \sigma}^{\nu \dagger}(\hat{\psi}_{\bm i \sigma}^{\nu })$ is the creation (annihilation) operator for an electron with spin $\sigma$ at the sublattice $\nu$ of the unit cell $\bm i$. $t$ parameterizes the hopping between adjacent sites, $\Delta_{\bm{i}}^\nu$ accounts for on-site s-wave pairing, $\mu$ is the electron chemical potential, and $V_i^\nu$ is the local on-site potential that is employed to capture the effect of disorder. $\hat n_{\bm i \sigma}^\nu = \hat \psi_{\bm i \sigma}^{\nu\dagger} \hat \psi_{\bm i \sigma}^\nu$ is the particle number operator at the site belonging to sublattice $\nu$ in unit cell $\bm i$.  Here after we define Pauli matrices $\bm \sigma = (\sigma_x, \sigma_y, \sigma_z)^T$ in spin space, $\bm \tau = (\tau_x, \tau_y, \tau_z)^T$ in particle-hole space and $\rho = (\rho_x, \rho_y, \rho_z)^T$ in sublattices space.

\begin{figure}[tb]
	\begin{center}
		\includegraphics[width=85mm]{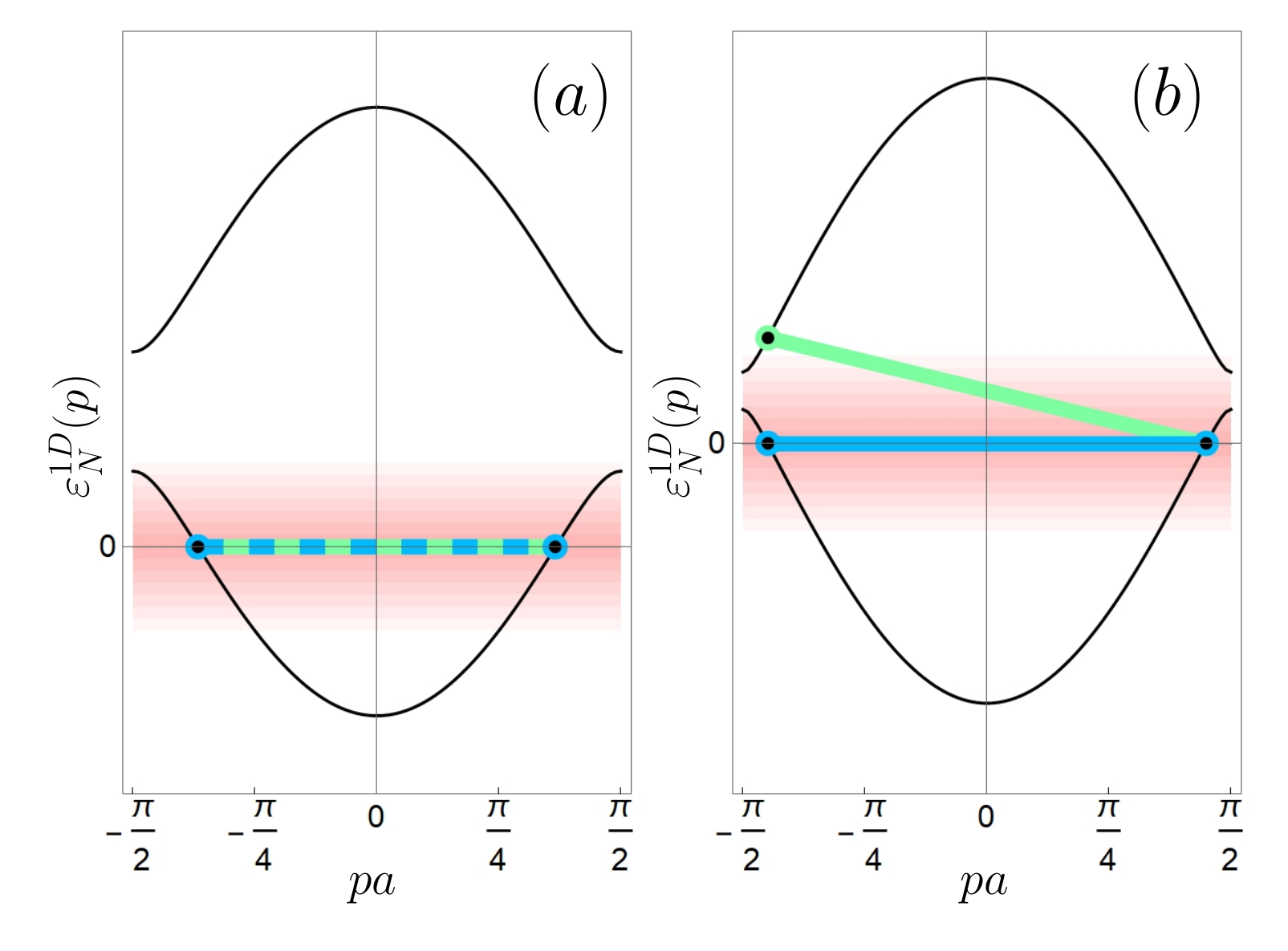}
		\caption{ Electron dispersion of the normal-state S in the reduced Brillouin zone (BZ) $pa \in [-\pi/2, \pi/2]$. For simplicity of visualization, a 1D system dispersion  $\varepsilon_N^{1D} = \mp \sqrt{\xi^2(p)+h^2} - \mu$ is demonstrated instead of a real dispersion $\varepsilon_N^{3D}$, considered in the paper. (a) Large $\mu$. The electronic states in the vicinity of the Fermi surface $\varepsilon_N^{1D} = 0$ allowed for pairing (pink region) do not involve the second electronic branch. Only $(\bm p, -\bm p)$ intraband singlet and intraband N\'eel triplet pairs are present (dashed blue-green). (b) Small $\mu$. Electronic states belonging to both branches are present in the vicinity of the Fermi surface and are allowed for pairing. Both intraband singlet (blue) and interband N\'eel triplet pairs (green) exist.}
        \label{fig:FS}
	\end{center}
\end{figure}

The Matsubara Green's function in the two-sublattice formalism is an $8 \times 8$ matrix in the direct product of spin, particle-hole and sublattice spaces. Introducing the two-sublattice Nambu spinor $\check \psi_{\bm i} = (\hat \psi_{{\bm i},\uparrow}^A, \hat \psi_{\bm i,\downarrow}^A, \hat \psi_{\bm i,\uparrow}^B,\hat \psi_{\bm i,\downarrow}^B, \hat \psi_{\bm i,\uparrow}^{A\dagger}, \hat \psi_{\bm i,\downarrow}^{A\dagger}, \hat \psi_{\bm i,\uparrow}^{B\dagger}, \hat \psi_{\bm i,\downarrow}^{B\dagger})^T$, we define the Green's function as follows: 
\begin{eqnarray}
\check G_{\bm i \bm j}(\tau_1, \tau_2) = -\langle T_\tau \check \psi_{\bm i}(\tau_1) \check \psi_{\bm j}^\dagger(\tau_2) \rangle,
\label{Green_Gorkov}
\end{eqnarray}
where $\langle T_\tau ... \rangle$ means  imaginary time-ordered thermal averaging. Further we consider the Green's function in the mixed representation:
\begin{eqnarray}
\check G(\bm R, \bm p) = F(\check G_{\bm i \bm j}) = \int d^3 r e^{-i \bm p(\bm i - \bm j)}\check G_{\bm i \bm j},
\label{mixed}
\end{eqnarray}
where $\bm R=(\bm i+\bm j)/2$ and the integration is over $\bm i - \bm j$. Then to make the resulting Gor'kov equations simpler it is convenient to define the following  transformed Green's function:
\begin{eqnarray}
\check {\tilde G}(\bm R, \bm p) = 
\left(
\begin{array}{cc}
1 & 0 \\
0 & -i\sigma_y
\end{array}
\right)_{\tau_{eh}} \rho_x e^{\frac{-\displaystyle ip_z a_z \rho_z}{\displaystyle 2}} \check G(\bm R, \bm p) \times \nonumber \\
e^{ \frac{\displaystyle ip_z a_z \rho_z}{\displaystyle 2}}
\left(
\begin{array}{cc}
1 & 0 \\
0 & -i\sigma_y
\end{array}
\right)_{\tau_{eh}} ,
\label{unitary}
\end{eqnarray}
where the subscript $\tau_{eh}$ means that the explicit matrix structure corresponds to the particle-hole space. The size of the two-site unit cell in the $z$-direction equals $2a_z$. The Gor'kov equation for $\check {\tilde G}(\bm R, \bm p)$ was derived in Ref.~\onlinecite{Bobkov2022_Neel} and takes the form:
\begin{align}
    [\check H(\bm R) \rho_x - \check \Sigma_{\mathrm{imp}}(\bm R) + \check T(\bm R, \bm p)] \check {\tilde G}(\bm R, \bm p) = 1,
    \label{gorkov_left} 
\end{align}

\begin{align*}
    \check H (\bm R) = i \omega_m \tau_z \rho_0 + \mu \tau_0 \rho_0 + \tau_z \check \Delta(\bm R) \rho_0 - \bm h (\bm R) \bm \sigma  \tau_z \rho_z ,
\end{align*}

\begin{eqnarray}
\check \Sigma_{\mathrm{imp}}  = \frac{1}{\pi N_F \tau_s}\int \frac{d^3p}{(2\pi)^3}[\rho_+ \check {\tilde G} \rho_+ +  \rho_- \check {\tilde G} \rho_-],
\end{eqnarray}

\begin{align*}
    \check T(\bm R)\check {\tilde G}(\bm R, \bm p) = t \frac{\rho_0+\rho_z}{2}\sum \limits_{\bm a}e^{-i \bm p \bm a} \check {\tilde G}(\bm R+ \frac{\bm a}{2}-\frac{\bm a_z}{2},\bm p)   \nonumber \\
+ t \frac{\rho_0-\rho_z}{2}\sum \limits_{\bm a}e^{-i \bm p \bm a} \check {\tilde G}(\bm R+ \frac{\bm a}{2}+\frac{\bm a_z}{2},\bm p),
\end{align*}
where $\rho_0 (\tau_0)$ is the unit matrix in the sublattice (particle-hole) space, $\rho_\pm = (\rho_x \pm i\rho_y)/2$, $\check \Delta (\bm R) =  \Delta(\bm R)\tau_+ + \Delta^*(\bm R) \tau_-$ with $\tau_\pm = (\tau_x \pm i \tau_y)/2$, $\omega_m = \pi T(2m+1)$ is the fermionic Matsubara frequency, $\tau_s$ is the quasiparticle mean free time at nonmagnetic impurities, and $N_F$ is the momentum-averaged density of states (DOS) at the Fermi surface. $\bm a =\pm\bm a_{x,y,z} $; see Fig.~\ref{fig:setup} for a visual definition of these lattice vectors. 

In this work, we assume that the system is spatially homogeneous along the S/AF interface, that is, the Green's function does not depend on in-plane coordinates $y$ and $z$. In addition, in the case of a thin S layer with $d_S \ll \xi_S$, where $d_S$ is its thickness along the $x$-direction and $\xi_S$ is the superconducting coherence length, the Green's function $\check {\tilde G}(\bm R, \bm p)$ practically does not depend on the $x$-coordinate, and Eq.~(\ref{gorkov_left}) can be averaged over the thickness of the superconducting layer. Then the effective exchange field does not depend on spatial coordinates and takes the form $\bm h=\frac{\bm h_0 a_x}{d_S}$. Therefore, mathematically our thin-film S/AF bilayer is equivalent to a homogeneous superconductor in the presence of the N\'eel exchange field and the Gor'kov equation transforms to the effectively homogeneous equation: 

\begin{align}
    [\check H \rho_x  -\xi(\bm p) - \check \Sigma_{\mathrm{imp}}]\check {\tilde G}=1,
    \label{eq:gorkov_hom} 
\end{align}
where $\xi(\bm p) = -2t(\cos p_x a_x + \cos p_y a_y + \cos p_z a_z)$. The solution of Eq.~(\ref{eq:gorkov_hom}) in the clean case $\tau_s^{-1} = 0$, linearized with respect to $\Delta$, takes the form:
\begin{align}
    \check {\tilde G}=&g_0^{on}\tau_0 \rho_x + g_z^{on}\tau_z \rho_x + g_0^{AB}\tau_0 \rho_0 + g_z^{AB}\tau_z \rho_0 + \nonumber \\
    &\bm g_z^{on} \bm \sigma \tau_z \rho_y + \bm g_0^{on} \bm \sigma \tau_0 \rho_y + \nonumber \\
    &f_s^{on} \tau_y \rho_x + f_s^{AB} \tau_y \rho_0 + \bm f_t^{on} \bm \sigma \tau_y \rho_y + \bm f_t^{AB} \bm \sigma \tau_x \rho_z,
    \label{eq:gorkov_sol} 
\end{align}
where the first two lines represent the normal components of the Green's function and the last line describes the anomalous components of the Green's function, responsible for superconducting pairing. The components marked by the superscript $on$ correspond to on-site correlations. $\rho_x$-components of the on-site correlations describe conventional correlations equal for A and B sublattices, while  $\rho_y$-components correspond to staggered correlations flipping sign between A and B sublattices (N\'eel-type). The components marked by the superscript $AB$ correspond to the correlations between electrons belonging to different sublattices. $\rho_0$-components are even with respect to the interchange of the sublattice indices, and  $\rho_z$-components are odd. All the vector components in spin space $\bm g_{z,0}^{on}$, $\bm f_t^{on,AB}$ are aligned with $\bm h$. We do not write out explicit expressions for the normal components of the Green's function because they are quite lengthy and are not very important for our analysis. The anomalous components, accounting for the singlet correlations, take the form:
\begin{align}
    f_s^{on} &= -D^{-1}i \Delta (\omega_m^2 + \xi^2 + \mu^2 - h^2), \nonumber \\
    f_s^{AB} &= -2 D^{-1}i \Delta \mu \xi, 
    \label{eq:gorkov_anomalous_singlet} 
\end{align}
where $h=|\bm h|$ and 
\begin{align}
    D = &(h^2-\mu^2)^2+2 (h^2-\mu^2) \xi^2 + \nonumber \\
    &(\xi^2 + \omega_m^2)^2 + 2 (h^2+\mu^2) \omega_m^2 .
    \label{eq:D} 
\end{align}
It is seen that the singlet correlations are even-momentum and even-frequency correlations. The triplet correlations take the form:
\begin{align}
    \bm f_t^{on} &= 2 D^{-1}i \Delta \omega_m \bm h, \nonumber \\
    \bm f_t^{AB} &= 2 D^{-1} \Delta \xi \bm h, 
    \label{eq:gorkov_anomalous_triplet} 
\end{align}
The on-site triplet correlations are even-momentum and odd-frequency. The $AB$-triplets are even-momentum, even-frequency, but odd with respect to the sublattice index correlations.

In the presence of impurities the general structure of the superconducting correlations described above remains unchanged.

The superconducting order parameter $\Delta$ in S is calculated self-consistently:

\begin{align}
    \Delta=-T\sum\limits_{|\omega_m|<\Omega_D} g\int \frac{\mbox{Tr}(\check {\tilde G}(\bm p) \tau_+\sigma_0\rho_x)}{8} \frac{d^3p}{(2\pi)^3},
    \label{eq:self-consistency}
\end{align}
where $g$ is the pairing constant and $\Omega_D$ is the Debye frequency.

In Refs.~\onlinecite{Bobkov2022_Neel,Fyhn2022} Eq.~(\ref{gorkov_left}) was used as a starting point for derivation of two different quasiclassical formalisms for description of S/AF heterostructures. The quasiclassical formalism is a powerful method for treating nonhomogeneous problems in superconducting mesoscopics. However, here we cannot use the quasiclassics because it does not allow for studying the crossover between the opposite regimes of the superconductivity suppression in S/AF hybrids: the N\'eel dominated suppression and the disorder-dominated suppression. The point is that the corresponding quasiclassical theories \cite{Bobkov2022_Neel,Fyhn2022} work in different parameter regions and none of them can restore the results of the other. 

The main idea behind quasiclassical theory is that most of the interesting physics happens close to the Fermi surface. Therefore, it is of interest to isolate the contribution from states close to the Fermi surface. In AFs due to the doubling of the unit cell, two bands of electrons spectra in the reduced Brillouin zone arise. In the normal state, they take the form $\varepsilon_N^{3D}(\bm p) = -\mu \pm \sqrt{\xi(\bm p)^2+h^2}$. 
The quasiclassical theory  developed in Ref.~\onlinecite{Fyhn2022} is valid if only states belonging to one of the bands are present in the vicinity of the Fermi surface. It is realized for large values of the chemical potential $\mu$, see Fig.~\ref{fig:FS}(a). In the framework of this quasiclassical theory all interband effects are  lost. On the other hand, the theory developed in Ref.~\onlinecite{Bobkov2022_Neel} assumes that the Fermi level is close to the energy of the bands crossing, see Fig.~\ref{fig:FS}(b). This fact allows for taking into account interband effects in the superconductor but imposed strong restrictions on value of chemical potential in it, which should be small. 

The nonquasiclassical approach in terms of the Gor'kov Green's functions, considered here, allows for obtaining the both limiting cases and studying the crossover between them.  

\section{Critical temperature of the thin-film S/AF bilayer for arbitrary impurity strength}
\label{results}

\begin{figure}[tb]
	\begin{center}
		\includegraphics[width=82mm]{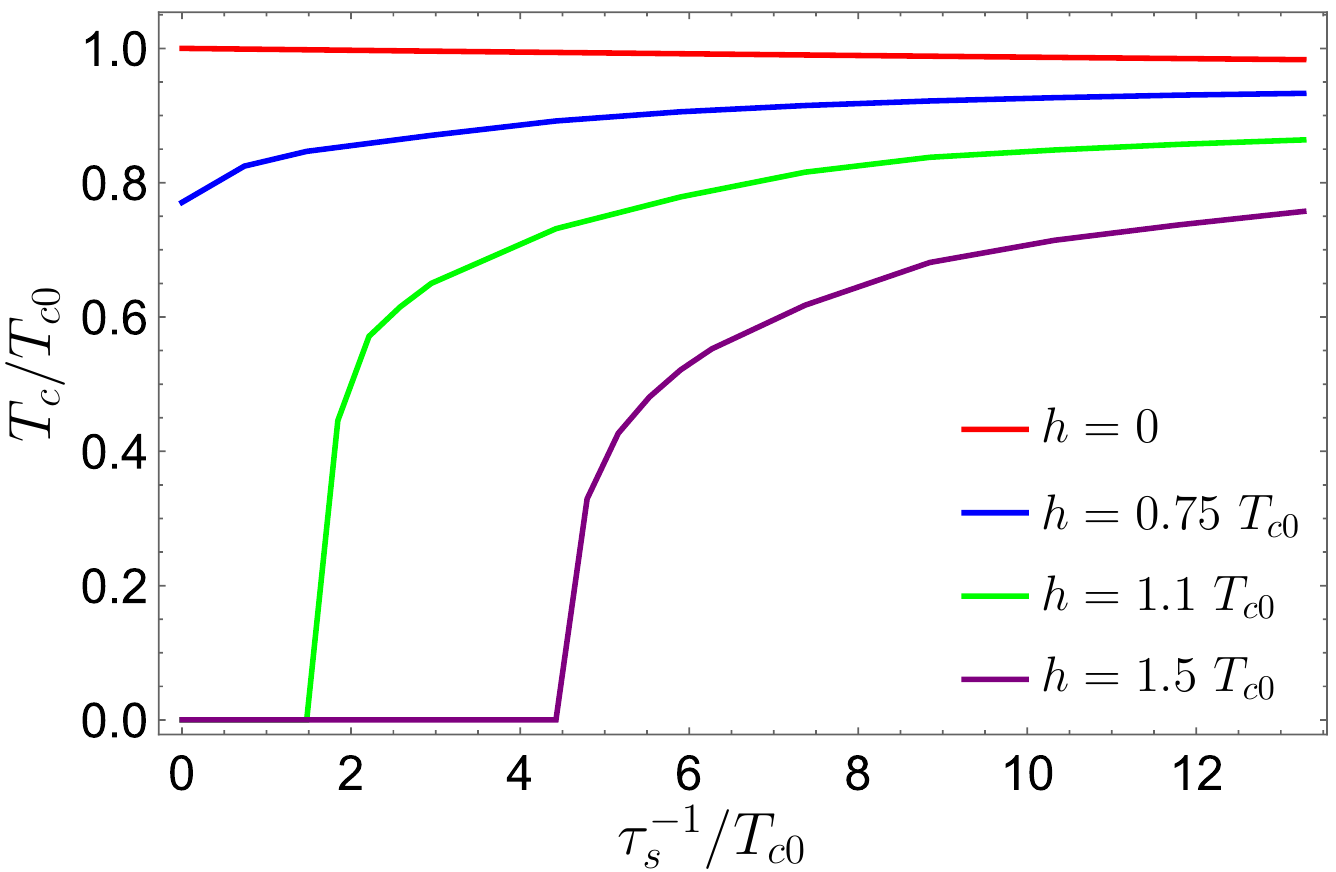}
		\caption{$T_c(\tau_s^{-1})$ at $\mu=0$ for different exchange fields $h$. $T_c$ is normalized to the value of the critical temperature of the isolated S film $T_{c0}=0.0067 t$.}
        \label{fig:fig_A}
	\end{center}
\end{figure}

\begin{figure}[tb]
	\begin{center}
		\includegraphics[width=85mm]{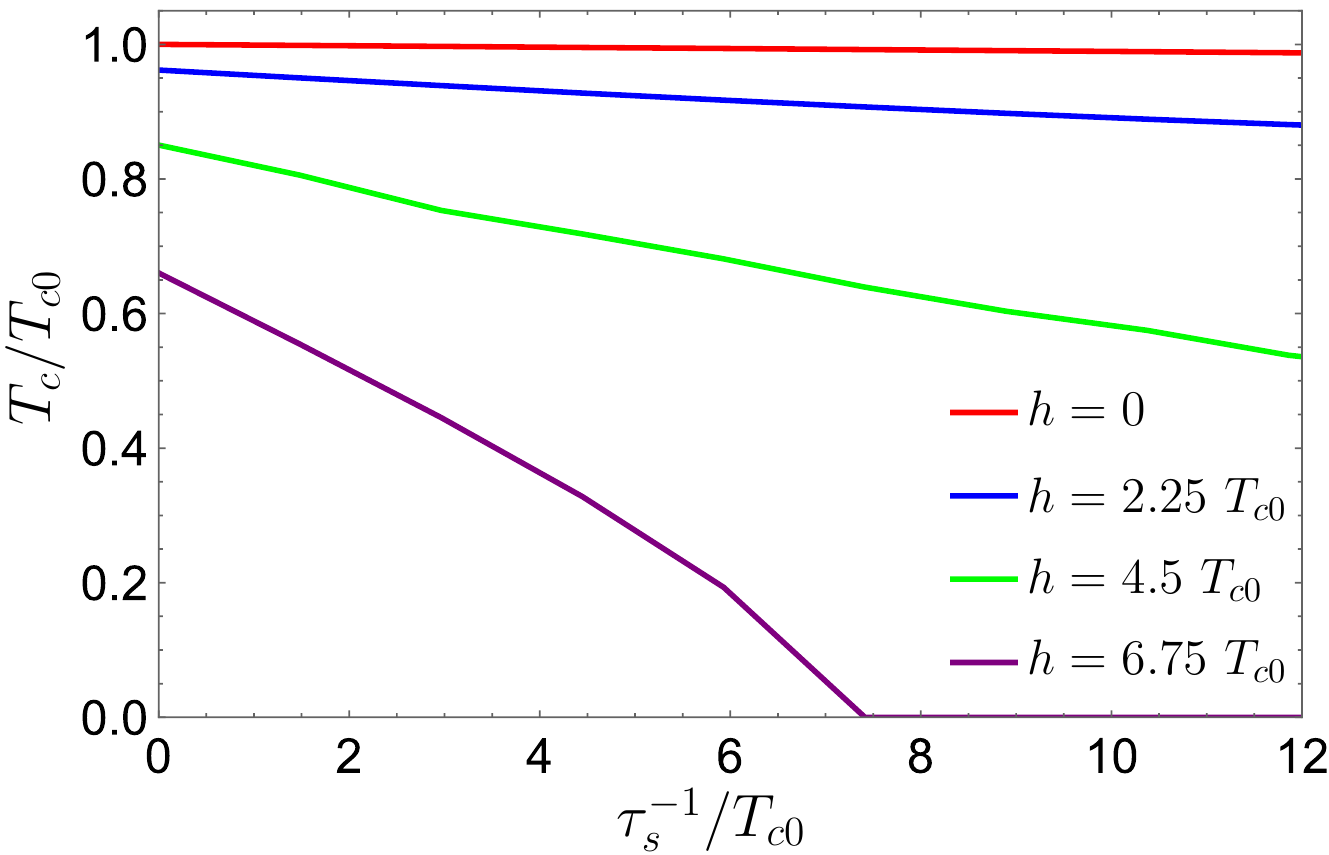}
		\caption{$T_c(\tau_s^{-1})$ at $\mu=0.1t$ for different exchange fields $h$.}
        \label{fig:fig_B}
	\end{center}
\end{figure}

\begin{figure}[tb]
	\begin{center}
		\includegraphics[width=85mm]{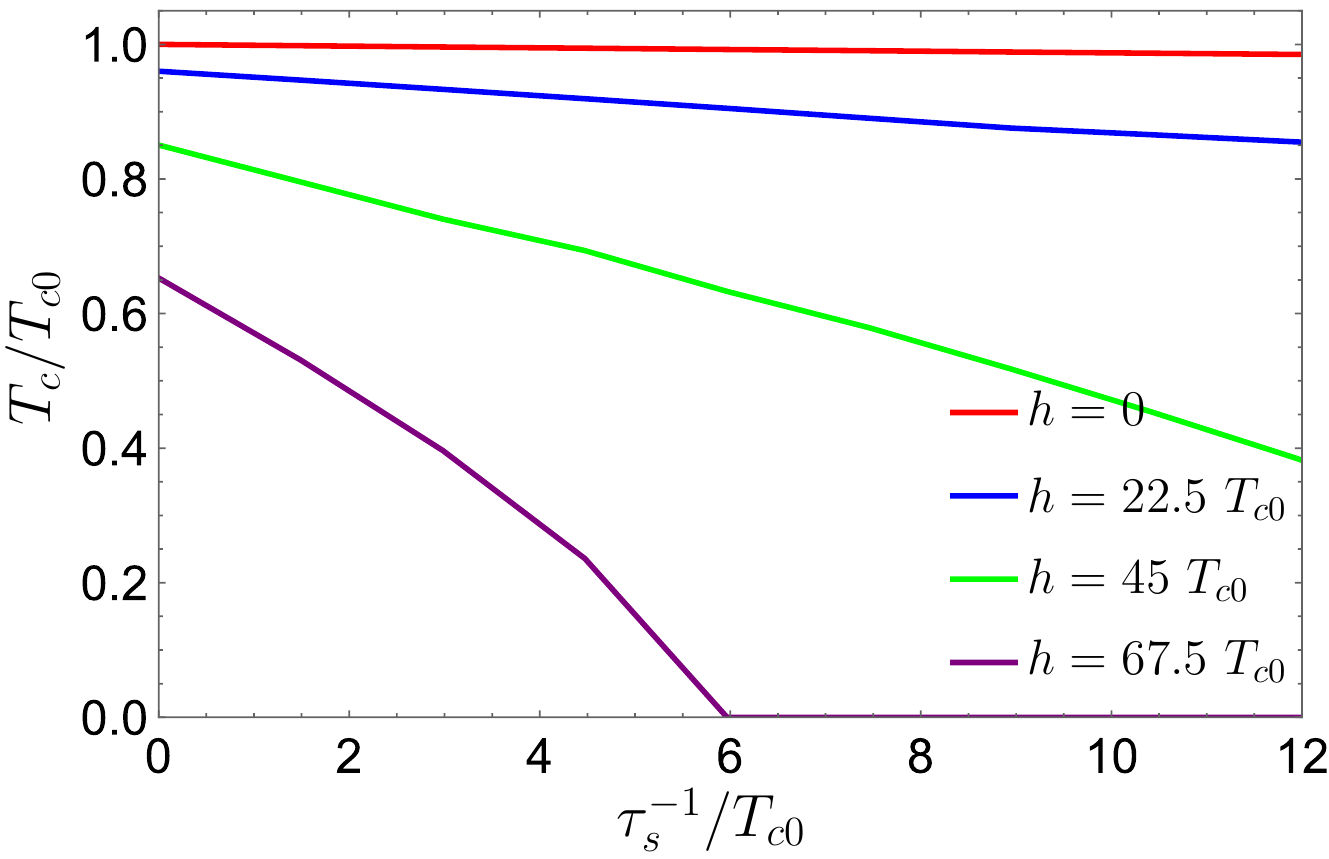}
		\caption{$T_c(\tau_s^{-1})$ at $\mu=t$ for different exchange fileds $h$.}
        \label{fig:fig_C}
	\end{center}
\end{figure}

Let us begin by considering the opposite limiting cases: N\'eel triplets dominated dependence of the critical temperature on impurity strength and disorder dominated behavior. Figs.~\ref{fig:fig_A}-\ref{fig:fig_C} demonstrate the critical temperature of the S/AF bilayer as a function of $\tau_s^{-1}$, which is obtained numerically from Eq.~(\ref{eq:self-consistency}). The results shown in Fig.~\ref{fig:fig_A} are calculated at $\mu=0$ and represent a typical example of the dependence in the regime when the main depairing mechanism  is the generation of the N\'eel triplets. It can be seen that at $h \neq 0$ the critical temperature grows with the disorder strength (blue curve) or even appears at some nonzero $\tau_s^{-1}$ (green and purple). This behavior is in agreement with predictions of Ref.~\onlinecite{Bobkov2022_Neel} and is explained as follows. There are N\'eel triplets in the system and they suppress the critical temperature of the singlet superconductivity. In the clean limit  $\tau_s^{-1}=0$ their amplitude is maximal. Due to the interband nature of the N\'eel pairing, see Fig.~\ref{fig:FS}(b) and Ref.~\onlinecite{Bobkov2022_Neel}, it is gradually reduced with impurity strength and, consequently, the critical temperature grows. 

Figs. \ref{fig:fig_B} and \ref{fig:fig_C} are evaluated at $\mu=0.1t$ and $\mu=t$, respectively. They represent the opposite limit when the dependence $T_c(\tau_s^{-1})$ is dominated by the impurity suppression. The both cases correspond to $\mu \gg T_{c0}$, where $T_{c0}=0.0067t$ is the critical temperature of the isolated S film. In this limit only one of two electronic bands, existing in the Brillouin zone, is in the vicinity of the Fermi level. This limit does not mean that the N\'eel triplets vanish.  As can be seen from Figs. \ref{fig:fig_B} and \ref{fig:fig_C}, in the clean limit (at $\tau_s^{-1} = 0$) the critical temperature is gradually suppressed with increasing $h$. It is the manifestation of the superconductivity suppression by the N\'eel triplets. By comparing Figs.~\ref{fig:fig_B} and \ref{fig:fig_C} one can notice that the degree of the $T_c$ suppression at $\tau_s^{-1} = 0$ is the same in these figures. At $(h,\mu) \gg T_{c0}$ it is proportional to the parameter $h/\mu$. This is because the amplitude of the N\'eel triplets is controlled by this parameter. However, at $\mu \gg T_{c0}$ the N\'eel triplets are mainly of intraband physical origin. We will get back to this issue below. For this reason they are practically not destroyed by the nonmagnetic impurities.  

Therefore, the impurity suppression plays the main role in the dependence of $T_c$ on the impuritiy strength in this limit. The mechanism of the superconductivity suppression in S/AF heterostructures  by nonmagnetic impurities was explained in detail in Ref.~\onlinecite{Fyhn2022_1}. The underlying physical reason is that the amplitude of wave function of electrons is different for A and B sublattices because for electron with spin up it is energetically favorable to be localized on B sublattice and for electron with spin down — on A sublattice. Therefore, for an electron the impurities become effectively magnetic. The parameter, controlling the strength of the impurity suppression is $(h/\mu)^2 \tau_s^{-1}$, as it was reported in Ref.~\onlinecite{Fyhn2022_1}.

\begin{figure}[tb]
	\begin{center}
		\includegraphics[width=88mm]{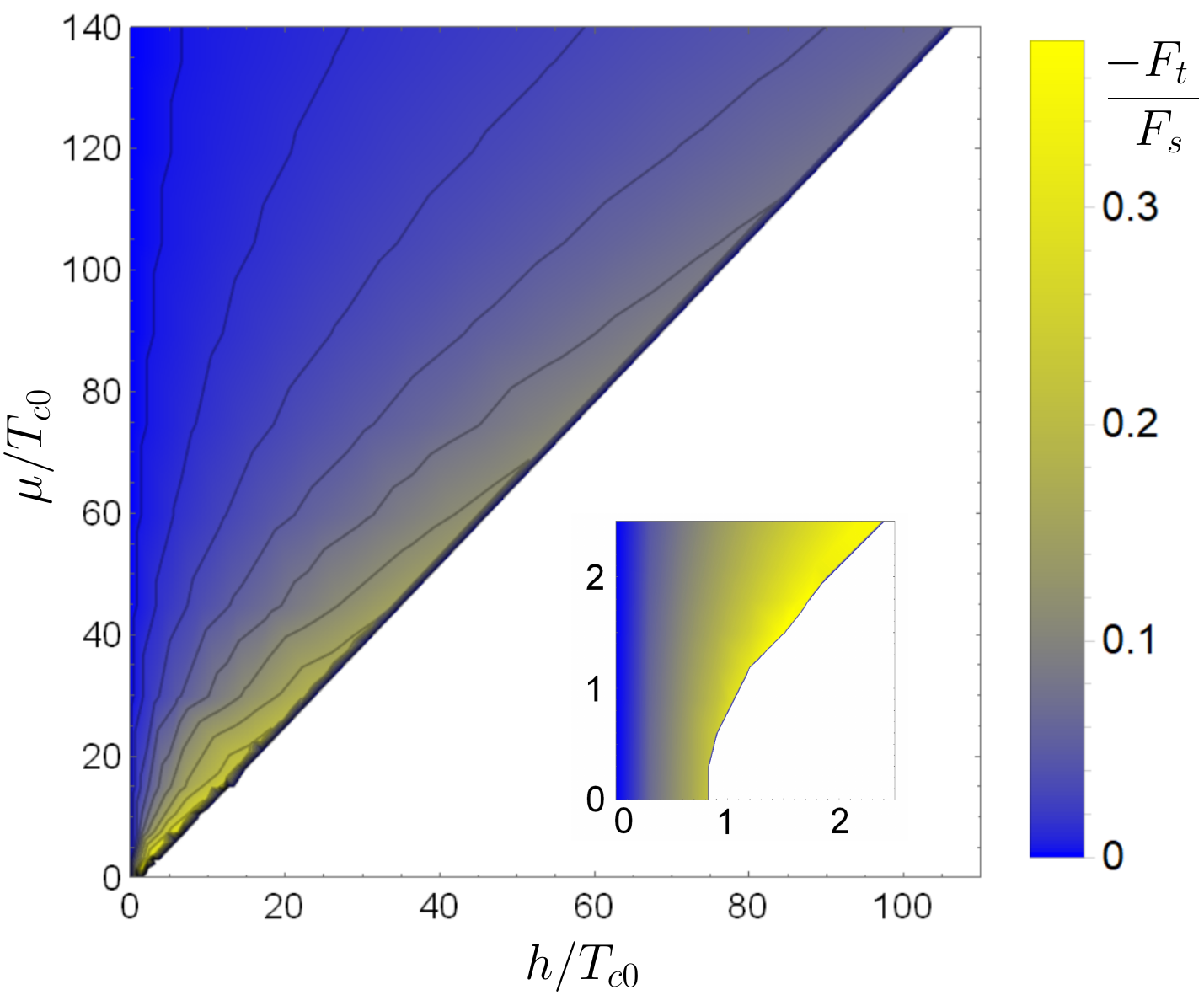}
		\caption{Amplitude of the triplet correlations relative to the singlet amplitude as function of $(h,\mu)$. Inset: region $(h,\mu) \sim T_{c0}$ on a larger scale. Clean limit $\tau_s^{-1} = 0$. $T \to T_c$.}
        \label{fig:triplets}
	\end{center}
\end{figure}

Now, after discussing the limiting cases let us consider the behavior of the S/AF system in the entire range of parameters. We begin by discussing the N\'eel triplets and the suppression of the critical temperature in the clean limit and then turn to the influence of impurities. Fig.~\ref{fig:triplets} represents the relative value of the N\'eel triplet correlations with respect to the value of the singlet correlations in the $(\mu,h)$-plane at $\tau_s^{-1}=0$ and near the critical temperature. The amplitude of the on-site triplet and singlet correlations are calculated as [we choose $\bm h = h \hat {\bm z}$] 
\begin{eqnarray}
F_t = T\sum_{\omega_m >0} \int \frac{{\rm Tr}[\check {\tilde G}(\bm p)\tau_+ \sigma_z \rho_y]}{8}\frac{d^3 p}{(2 \pi)^3},
\label{eq:triplets}
\end{eqnarray}
\begin{eqnarray}
F_s = T\sum_{\omega_m >0} \int \frac{{\rm Tr}[\check {\tilde G}(\bm p)\tau_+ \sigma_0 \rho_x]}{8}\frac{d^3 p}{(2 \pi)^3}.
\label{eq:singlets}
\end{eqnarray}
It is seen that the relative amplitude of the N\'eel triplet correlations reaches maximal value at the line $h/\mu = const \approx 0.8$. For larger values of this parameter the superconductivity in the system is fully suppressed. For a given exchange field the relative amplitude of the N\'eel triplets decreases with increasing $\mu$. However, we can see that the N\'eel triplet correlations can be essential even at large values of $\mu$. The interband pairing is not essential at $\mu \gg T_{c0}$. For this reason the N\'eel triplet correlations presented in Fig.~\ref{fig:triplets} are not associated  with the interband pairing (except for the region $(h,\mu) \sim T_{c0}$). The pairing correlations are completely intraband, as it is shown in Fig.~\ref{fig:FS}(a). However, the normal state eigenvectors of the electronic band crossed by the Fermi level represent a mixture of sign-preserving and sign-flipping components between the A and B sites of the same unit cell:
\begin{eqnarray}
\left( \begin{array}{c} \hat \psi_{\bm i \sigma}^A \\ \hat \psi_{\bm i \sigma}^B
\end{array}
\right)(\bm p) = \left[C_{p\sigma} \left( \begin{array}{c} 1 \\ 1
\end{array}
\right) + C_{f\sigma} \left( \begin{array}{c} 1 \\ -1
\end{array}
\right)\right]e^{i \bm p \bm i},
\label{eq:eigenvectors}
\end{eqnarray}
where
\begin{eqnarray}
C_{p(f)\sigma} = \frac{1}{2}\left[ \sqrt{1+\sigma h/\mu}\pm \sqrt{1-\sigma h/\mu} \right]
\label{eq:cpf}
\end{eqnarray}
are the sign-preserving (flipping) amplitudes of the eigenvector. Due to the presence of the sign-flipping component of the eigenvectors and its spin sensitivity the singlet homogeneous pairing between $\bm p$ and $-\bm p$ states at the Fermi level [see Fig.~\ref{fig:FS}(a)] is inevitably accompanied by the N\'eel sign-flipping triplet component. As it can be seen from Eq.~(\ref{eq:cpf}) the amplitude of the sign-flipping mixture is controlled by $h/\mu$ and is suppressed with growing of $\mu$, what is in agreement with the results of direct nonquasiclassical calculations, presented in Fig.~\ref{fig:triplets}.  

A deviation from this behavior is only observed at small $(\mu,h) \sim T_{c0}$. It is presented in the inset to Fig.~\ref{fig:triplets}. In this case there are two electronic branches in the vicinity of the Fermi level, which should be considered on equal footing, see Fig.~\ref{fig:FS}(b). The resulting N\'eel triplet pairing is mainly of the interband nature, what leads to its qualitatively different dependence on the impurity strength.

\begin{figure}[tb]
	\begin{center}
		\includegraphics[width=84mm]{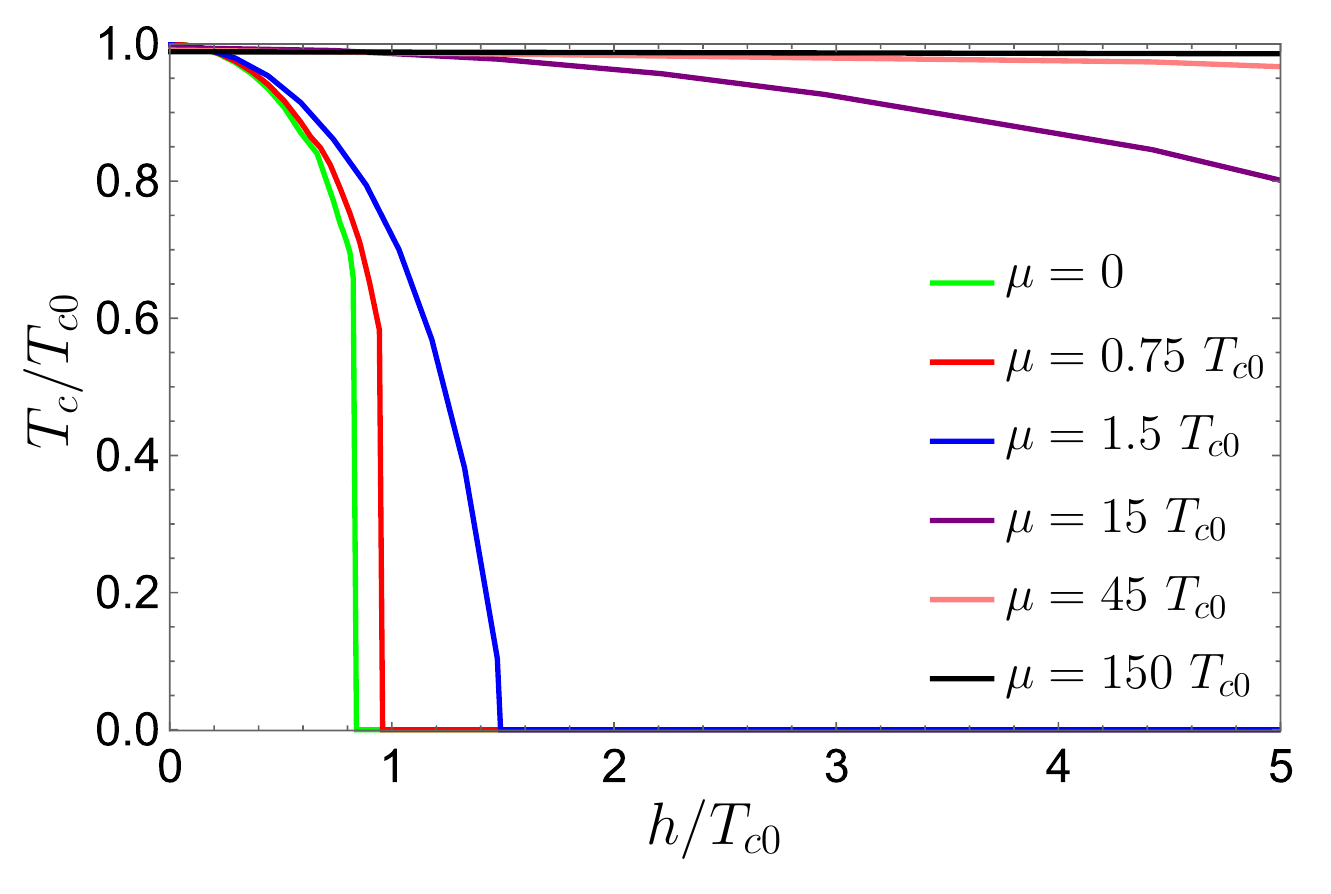}
		\caption{$T_c$ as a function of the N\'eel exchange field $h$ for different values of the chemical potential $\mu$. Clean case $\tau_s^{-1} = 0$.}
        \label{fig:fig_D}
	\end{center}
\end{figure}

\begin{figure}[tb]
	\begin{center}
		\includegraphics[width=85mm]{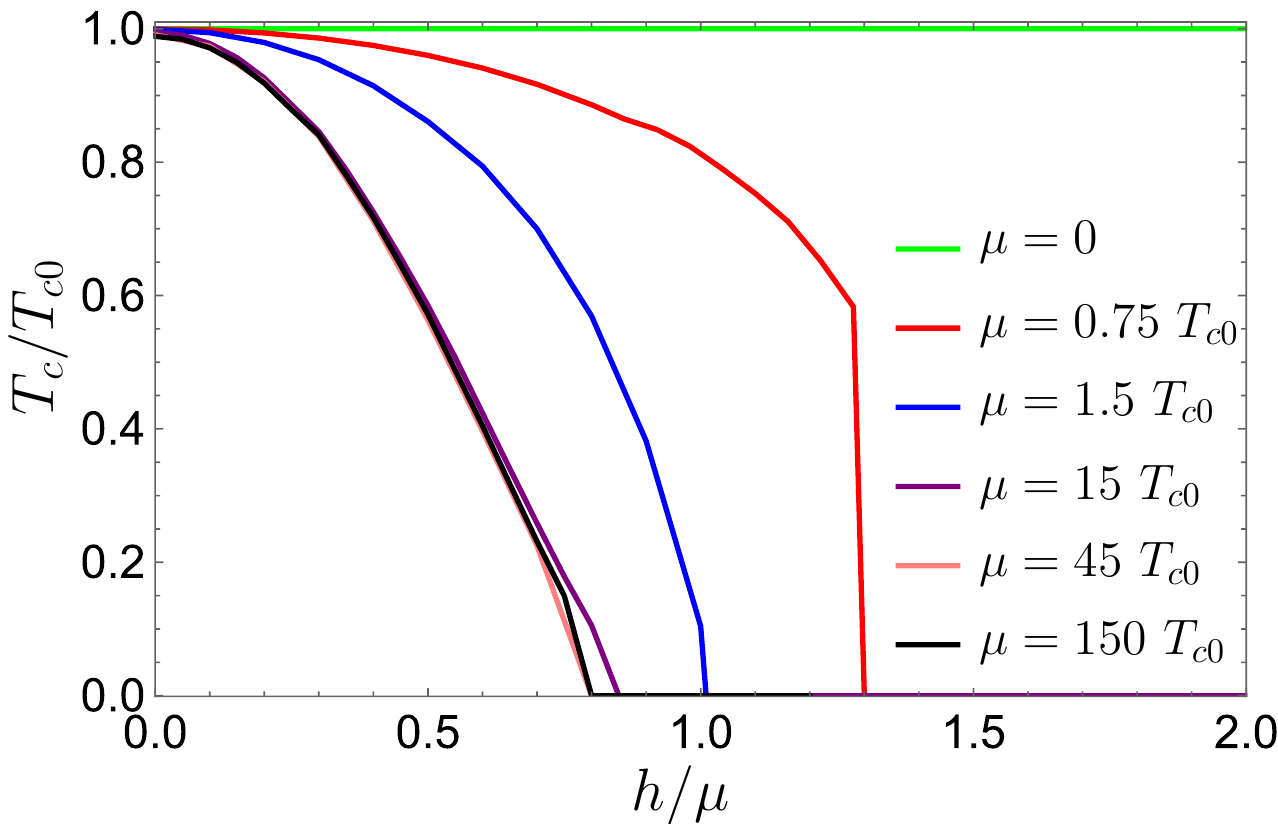}
		\caption{$T_c$ as a function of the ratio between the N\'eel exchange field $h$ and the chemical potential $\mu$ for different values of $\mu$. Clean case $\tau_s^{-1} = 0$. It is seen that at $\mu \gg T_{c0}$ the critical temperature depends only on the ratio $h/\mu$. The deviations from this law at $\mu \sim T_{c0}$ are also clearly seen. The green curve, corresponding to $\mu=0$, means that if $\mu$ and $h$ go to zero simultaneously keeping the fixed value of $h/\mu$, the critical temperature tends to $T_{c0}$.}
        \label{fig:fig_E}
	\end{center}
\end{figure}

Figs.~\ref{fig:fig_D}-\ref{fig:fig_E} show the details of the dependence $T_c(h,\mu)$ in the clean limit. Fig.~\ref{fig:fig_D} demonstrates that the suppression of the critical temperature by the N\'eel exchange field becomes weaker with increasing $\mu$. This is explained by the decrease of the N\'eel triplets for a given exchange field with increasing $\mu$. Fig.~\ref{fig:fig_E} represents the dependence of $T_c$ on the ratio $h/\mu$. It is seen that at large $\mu \gg T_{c0}$ the critical temperature depends only on the ratio $h/\mu$ and does not depend on these parameters separately. At $h>\mu$ the normal state of the superconducting layer becomes insulating because the Fermi level is located inside the antiferromagnetic gap. Therefore, there are no electrons near the Fermi level that can be paired. The superconductivity is fully suppressed even earlier, at $h/\mu \approx 0.8$. For this reason the N\'eel triplets do not exist for higher values of this ratio in Fig.~\ref{fig:triplets}. However, for small values of $\mu \sim T_{c0}$ and $h \sim T_{c0}$ assertion that the critical temperature depends only on the ratio $h/\mu$ is violated. Superconductivity exists even at $h>\mu$  despite the fact that the Fermi level is inside the antiferromagnetic gap \cite{Bobkov2022_Neel}. This is because  the gap itself is small and there are electrons beyond the gap, which are available for pairing.

\begin{figure}[tb]
	\begin{center}
		\includegraphics[width=85mm]{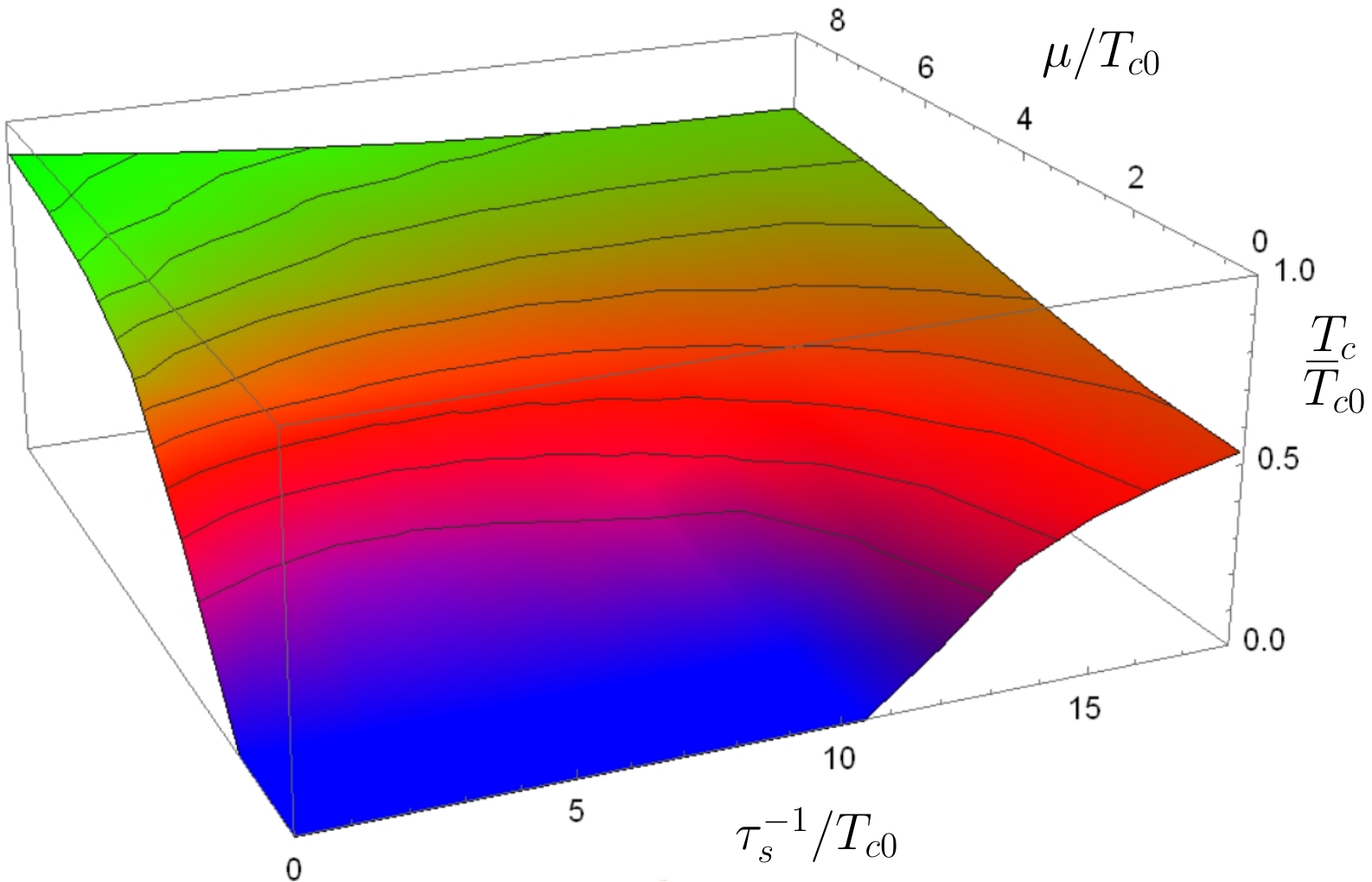}
		\caption{Dependence of the critical temperature on $(\tau_s^{-1},\mu)$ at $h=2.25 T_{c0}$.}
        \label{fig:impurities}
	\end{center}
\end{figure}

Now we discuss the effect of nonmagnetic impurities on the superconductivity of the S/AF hybrids in the entire range of parameters. Fig.~\ref{fig:impurities} represents the behavior of the critical temperature in the plane $(\tau^{-1}_s,\mu)$ for a given $h=2.25 T_{c0}$. Front and back edges of the image correspond to opposite limits. Front edge is the limit, where N\'eel triplets dominate and, consequently, superconductivity is restored with increase in impurity strength. Back edge represent the suppression of superconductivity by nonmagnetic impurities. The first regime is realized at small $\mu$ and the second one at large $\mu$, as it was discussed above. For intermediate $\mu$ we see the crossover between them. In particular, for a certain range of $\mu$ a nonmonotonic dependence $T_c(\tau^{-1}_s)$ is observed. The initial suppression of $T_c$ is changed by some growth. This is because the  singlet superconductivity is suppressed by the impurities more rapidly than the N\'eel triplets.

For larger and smaller values of $h$ the dependence $T_c(\tau^{-1}_s,\mu)$ is qualitatively the same. For larger $h$ the N\'eel triplets are stronger and, consequently, the region of suppressed superconductivity in the front left corner is wider. The suppression of superconductivity by the impurities is also stronger and occurs more rapidly because the strength of the suppression is $\sim (h/\mu)^2\tau^{-1}_s$. For weaker $h$ the degree of superconductivity suppression by the both factors is smaller.

\section{Conclusions}
\label{conclusions}
In this work possible mechanisms of suppression of superconductivity in thin-film S/AF bilayer structures with compensated antiferromagnets, namely, the N\'eel
triplets generation and the suppression by nonmagnetic impurities, are considered in the entire range of parameters $\tau^{-1}_s$, $\mu$ and $h$. In the clean limit $\tau^{-1}_s = 0$ and for small values of the chemical potential $\mu \lesssim T_{c0}$ the superconductivity is mainly suppressed by the interband N\'eel triplets. This leads to the growth of the critical temperature and even appearance of superconductivity with increasing the impurity strength $\tau^{-1}_s$. For large values of $\mu\gg T_{c0}$ the amplitude of the interband N\'eel triplets decreases and  their role as the superconductivity suppressing mechanism is reduced. The main suppressing mechanisms are intraband N\'eel triplets and scattering by nonmagnetic impurities, which behave as effectively magnetic at S/AF interfaces \cite{Buzdin1986,Fyhn2022_1} due to the spin-dependent atomic oscillations of the electron wave functions in the presence of the AF order. In this limit the superconducting critical temperature decreases with the impurity strength. For intermediate values of $\mu$ the crossover between the two limits is demonstrated. In particular, for a certain range of $\mu$ the both mechanisms contribute to the suppression on equal footing, what leads to the nonmonotonic dependence $T_c(\tau^{-1}_s)$. The theory developed in the work is based on exact Green's functions formalism and incorporates results of two different quasiclassical formalisms presented earlier in the literature\cite{Bobkov2022_Neel,Fyhn2022} as limiting cases. 

\section{Acknowledgments}
We thank Akashdeep Kamra for discussions. The financial support from the Russian
Science Foundation via the RSF project No.22-22-00522 is acknowledged. 

\bibliography{nonquasiclassical}

\end{document}